\documentclass[preprint]{emulateapj}
\usepackage{epsf}
\begin{document}

\submitted{The Astrophysical Journal, submitted}
\vspace{1mm}
\slugcomment{{\em The Astrophysical Journal, submitted}} 
%\shortauthors{KRAVTSOV, NAGAI \& VIKHLININ}
%\shorttitle{ON THE BARYON FRACTION IN GALAXY CLUSTERS}

\title{Effects of cooling and star formation\\ on the
baryon fractions in clusters}

\author{Andrey
V. Kravtsov\altaffilmark{1,2,3}, Daisuke Nagai\altaffilmark{1,3}, 
Alexey A. Vikhlinin\altaffilmark{4,5}}

\begin{abstract}  
  We study the effects of radiative cooling and galaxy formation on
  the baryon fractions in clusters using high-resolution cosmological
  simulations that resolve formation of cluster galaxies. The
  simulations of nine individual clusters spanning a decade in mass
  are performed with the shock-capturing eulerian adaptive mesh
  refinement $N$-body$+$gasdynamics ART code. For each cluster the
  simulations were done in the adiabatic regime (without dissipation)
  and with radiative cooling and several physical processes critical
  to various aspects of galaxy formation: star formation, metal
  enrichment and stellar feedback.  We show that radiative cooling of
  gas and associated star formation increase the total baryon
  fractions within radii as large as the virial radius.  The effect is
  strongest within cluster cores, where the simulations with cooling
  have baryon fractions larger than the universal value, in contrast
  to the adiabatic simulations in which the fraction of baryons is
  substantially smaller than universal. At larger radii ($r\gtrsim
  r_{500}$) the cumulative baryon fractions in simulations with
  cooling are close to the universal value. The {\it gas} fractions in
  simulations with dissipation are reduced by $\approx 20-40\%$ at
  $r<0.3r_{\rm vir}$ and $\approx 10\%$ at larger radii compared to
  the adiabatic runs, because a fraction of gas is converted into
  stars. There is an indication that gas fractions within different
  radii increase with increasing cluster mass as $f_{\rm gas}\propto
  M_{\rm vir}^{0.2}$.  We find that the total baryon fraction within
  the cluster virial radius does not evolve with time in both
  adiabatic simulations and in simulations with cooling. The gas
  fractions in the latter decrease slightly from $z=1$ to $z=0$ due to
  ongoing star formation.
  
  Finally, to evaluate systematic uncertainties in the baryon fraction
  in cosmological simulations we present a comparison of gas fractions
  in our adiabatic simulations to re-simulations of the same objects
  with the entropy-conserving SPH code Gadget. The cumulative gas
  fraction profiles in the two sets of simulations on average agree to
  better than $\approx 3\%$ outside the cluster core ($r/r_{\rm
    vir}\gtrsim 0.2$), but differ by up to 10\% at small radii.  The
  differences are smaller than those found in previous comparisons of
  eulerian and SPH simulations.  Nevertheless, they are systematic and
  have to be kept in mind when using gas fractions from cosmological
  simulations. 
\end{abstract}

% User-supplied List of keywords.


\altaffiltext{1}{Department of Astronomy and
Astrophysics, Kavli Institute for Cosmological Physics, 5640 South
Ellis Ave., The University of Chicago, Chicago, IL 60637}
\altaffiltext{2}{The Enrico Fermi Institute, The University of Chicago, Chicago, IL 60637}
\altaffiltext{3}{{\tt daisuke,andrey@oddjob.uchicago.edu}}
\altaffiltext{4}{Harvard-Smithsonian Center for Astrophysics, 60 Garden Street, Cambridge, MA 02138}
\altaffiltext{5}{Space Research Institute, 8432 Profsojuznaya St., GSP-7, Moscow 117997, Russia}

%----------------------
\section{Introduction}
\label{sec:intro}
%----------------------

In hierarchical models of structure formation clusters of galaxies
form via collapse of a large representative volume and are therefore
expected to contain the baryons and dark matter in proportions close
to the universal average. This fact together with independent
measurements of the universal baryon fraction can be used as a
powerful constraint on the mean density of matter in the universe
\citep[e.g.,][]{white_etal93,evrard97}. The apparent evolution of
baryon fraction with redshift can be used to put constraints on the
energy density content of the universe and properties of dark energy
\citep{sasaki96,pen97}. The universality of cluster baryon fractions
can be used to estimate total cluster mass and construct mass
functions of clusters at different redshifts \citep{vikhlinin_etal03}.

Recently, deep X-ray observations of nearby and high-redshift clusters
have been used to measure the cluster baryon fraction and constrain
the power spectrum shape and normalization
\citep{voevodkin_vikhlinin04}, the matter and dark energy density of
the universe
\citep{allen_etal02,allen_etal04,vikhlinin_etal03,chen_ratra04}, and
even the neutrino mass and the equation of state of dark energy
\citep{allen_etal04}. Sunyaev-Zeldovich (SZ) effect observations
combined with X-ray observations have also been used to measure the
cluster baryon fraction and constrain the matter energy density of the
universe \citep{grego_etal01}. These measurements are complementary to
(and competitive with) cosmological constraints using supernovae type
Ia and anisotropies of the cosmic microwave background and are likely
to be improved. However, they rely on the key assumptions that the
baryon fraction within a given radius can be measured reliably, is
independent of redshift, and can be converted to the universal
fraction by the correction factor derived from cosmological
simulations.

In order to improve the baryon fraction based cosmological constraints
and to better understand associated systematic uncertainties, these
assumptions have to be carefully tested with simulations. During the
last decade, a number of studies have considered the cluster baryon fraction
and its evolution in cosmological simulations in the adiabatic
regime (i.e., without radiative cooling).  The main results of these
studies are: (1) the baryon fraction within the cluster virial radius
evolves only very weakly with time and (2) is in general $\approx
10-15\%$ lower than the universal average
\citep{evrard90,metzler_evrard94,navarro_etal95,lubin_etal96,eke_etal98,frenk_etal99,mohr_etal99,bialek_etal01}; 
(3) this ``baryon depletion'' can be enhanced by strong pre-heating of
gas \citep{bialek_etal01,borgani_etal02,muanwong_etal02,kay_etal03}.
Note, however, that high-resolution eulerian simulations of
\citet{anninos_norman96} exhibit baryon fraction larger than
universal. Moreover, systematic comparison of adiabatic simulations of
a massive cluster performed with different numerical codes
\citep{frenk_etal99} has revealed systematic differences.  The
eulerian shock-capturing codes have generally produced clusters with
little or no ``baryon depletion,'' while gas fractions in SPH
simulations were $\approx 10\%$ below the universal value. This
indicates that there are systematic differences in gas and dark matter
density profiles in simulations done with different numerical schemes.

More recently, several groups have presented analysis of baryon
fractions in simulations that include gas cooling and star formation
\citep{muanwong_etal02,kay_etal03,valdarnini03,ettori_etal04}.  These
simulations have shown that cooling can increase the total baryon
fraction (gas + stars) and change its dependence on cluster mass and
cluster-centric radius compared to the adiabatic simulations. 

The purpose of the present paper is twofold. First, given the
systematic differences between numerical codes identified in
\citet{frenk_etal99}, we would like to present a systematic study of
the cluster baryon fractions in eulerian adaptive mesh refinement
(AMR) simulations and compare them to the results obtained with 
modern SPH codes.  Second, we investigate the effects of radiative
cooling and star formation on the baryon fractions in the very
high-resolution AMR simulations that resolve formation of individual
cluster galaxies.

The paper is organized as follows. In \S~\ref{sec:sim} we describe our
simulations and implementation of radiative cooling and star
formation.  In \S~\ref{sec:artvsgadget} we compare gas fraction
profiles in our adiabatic simulations with the simulations of the same
clusters performed with the entropy-conserving Gadget code. In
\S~\ref{sec:resolution} we discuss convergence tests and assess
effects of resolution.  In \S~\ref{sec:results} we present our main
results and compare them to the previous results in
\S~\ref{sec:previous}. Finally, in \S~\ref{sec:discussion} we discuss
our conclusions and implications of the presented results for cluster
observations and their use as cosmological probes.

% CL1 = Coma-size;
% CL2 = CL5
% CL3 = CL6
% CL4 = CL7
% CL5 = CL9
% CL6 = CL10
% CL7 = CL11
% CL8 = CL14
% CL9 = CL24
% 
%
\begin{deluxetable*}{cccccccccc}
%\tablenum{1}
\tablecaption{Simulated cluster sample at $z=0$}
\tablehead{
\\
\multicolumn{1}{c}{Name}&
\multicolumn{1}{c}{$R_{\rm vir}$}&
\multicolumn{1}{c}{$M_{\rm vir}$}  &
\multicolumn{1}{c}{$M_{\rm 180m}$}  &
\multicolumn{1}{c}{$M_{\rm 200c}$}  &
\multicolumn{1}{c}{$M_{\rm 500c}$} &
\multicolumn{1}{c}{$M_{\rm 2500c}$} &
\multicolumn{1}{c}{$\langle T^{\rm ad}_{\rm X}\rangle$} &
\multicolumn{1}{c}{$\langle T^{\rm csf}_{\rm X}\rangle$} &
\\
\multicolumn{1}{c}{}&
\multicolumn{1}{c}{$h^{-1}\rm Mpc$}&
\multicolumn{1}{c}{}&
\multicolumn{1}{c}{} & 
\multicolumn{1}{c}{$10^{14}h^{-1}\rm\ M_{\odot}$}&
\multicolumn{1}{c}{} & 
\multicolumn{1}{c}{} & 
\multicolumn{1}{c}{keV} &
\multicolumn{1}{c}{keV} &
\\
}
\startdata
\\
CL1 & 1.911 & 8.21 & 9.62 & 6.73 & 5.31 & 2.78 & 6.9 & 8.6 \\
CL2 & 1.205 & 2.06 & 2.32 & 1.75 & 1.34 & 0.75 & 2.7 & 2.6 \\
CL3 & 1.227 & 2.17 & 2.45 & 1.87 & 1.32 & 0.69 & 2.7 & 3.8 \\
CL4 & 1.187 & 1.97 & 2.37 & 1.65 & 1.24 & 0.63 & 2.8 & 3.2 \\
CL5 & 1.016 & 1.23 & 1.41 & 1.01 & 0.78 & 0.39 & 1.8 & 1.7 \\
CL6 & 0.906 & 0.87 & 0.99 & 0.78 & 0.62 & 0.34 & 1.6 & 2.5 \\
CL7 & 0.947 & 1.00 & 1.22 & 0.66 & 0.48 & 0.24 & 1.4 & 1.0 \\
CL8 & 0.966 & 1.06 & 1.24 & 0.88 & 0.65 & 0.34 & 1.7 & 2.1 \\
CL9 & 0.795 & 0.59 & 0.68 & 0.51 & 0.35 & 0.13 & 0.8 & 1.2 \\
\enddata
%\normalsize
\label{tab:sim}
\end{deluxetable*}

%----------------------
\section{Simulations}
\label{sec:sim}
%----------------------

In this study, we analyze high-resolution cosmological simulations of
nine cluster-size systems in the ``concordance'' flat {$\Lambda$}CDM
model: $\Omega_{\rm m}=1-\Omega_{\Lambda}=0.3$, $\Omega_{\rm
b}=0.04286$, $h=0.7$ and $\sigma_8=0.9$, where the Hubble constant is
defined as $100h{\ \rm km\ s^{-1}\ Mpc^{-1}}$, and $\sigma_8$ is the
power spectrum normalization on $8h^{-1}$~Mpc scale.  The simulations
were done with the Adaptive Refinement Tree (ART)
$N$-body$+$gasdynamics code \citep{kravtsov99, kravtsov_etal02}, a
Eulerian code that uses adaptive refinement in space and time, and
(non-adaptive) refinement in mass \citep{klypin_etal01} to reach the
high dynamic range required to resolve cores of halos formed in
self-consistent cosmological simulations.

To set up initial conditions we first ran a low resolution simulation
of $80h^{-1}$~Mpc and $120h^{-1}$~Mpc boxes and selected nine clusters
with the virial masses ranging from $\approx 7\times10^{13}$ to
$8.2\times 10^{14}h^{-1}{\ \rm M_{\odot}}$. The properties of clusters
at the present epoch are listed in Table~\ref{tab:sim}. The
perturbation modes in the lagrangian region corresponding to the
sphere of several virial radii around each cluster at $z=0$ was then
re-sampled at the initial redshift, $z_i=49$ for eight clusters and
$z_i=25$ for the most massive cluster in the sample.  For the
Coma-size cluster we have resampled radius of $1.5R_{\rm vir}(z=0)$, 
while for the rest of the clusters the resampling sphere had
radius of $5R_{\rm vir}$. During the resampling we retained the
previous large-scale waves intact but included additional small-scale
waves, as described by \citet{klypin_etal01}.  The resampled
lagrangian region of each cluster was then re-simulated with high
dynamic range.

High-resolution simulations were run using 128$^3$ uniform grid and 8
levels of mesh refinement in the computational boxes of $80h^{-1}$~Mpc
for CL2-CL9 and $120h^{-1}$~Mpc for the Coma-size CL1, which
corresponds to the dynamic range of $128\times 2^8=32768$ and peak
formal resolution of $80/32,768\approx 2.44h^{-1}\ \rm kpc$,
corresponding to the actual resolution of $\approx 2\times 2.44\approx
5h^{-1}\ \rm kpc$. Only the region of $\sim 3-10h^{-1}\ \rm Mpc$
around the cluster was adaptively refined, the rest of the volume was
followed on the uniform $128^3$ grid. The mass resolution corresponds
to the effective $512^3$ particles in the entire box, or the Nyquist
wavelength of $\lambda_{\rm Ny}=0.469h^{-1}$ and $0.312h^{-1}$ {\it
  comoving} megaparsec for CL1 and CL2-9, respectively, or
$0.018h^{-1}$ and $0.006h^{-1}$ Mpc in the physical units at the
initial redshift of the simulations. The dark matter particle mass in
the region around the cluster was $2.7\times 10^{8}h^{-1}{\rm\ 
  M_{\odot}}$ for CL2-CL9 and $9.1\times 10^{8}h^{-1}{\rm\ M_{\odot}}$
for CL1, while other regions were simulated with lower mass
resolution.

As the zeroth-level fixed grid consisted of only $128^3$ cells, we
started the simulation already pre-refined to the 2nd level
($l=0,1,2$) in the high-resolution lagrangian regions of clusters.
This is done to ensure that the cell size is equal to the mean
interparticle separation and all fluctuations present in the initial
conditions are evolved properly. During the simulation, the
refinements were allowed to the maximum $l=8$ level and refinement
criteria were based on the local mass of DM and gas in each cell. The
logic is to keep the mass per cell approximately constant so that the
refinements are introduced to follow the collapse of matter in a
quasi-lagrangian fashion. For the DM, we refine the cell if it
contains more than two dark matter particles of the highest mass
resolution specie.  For gas, we allow the mesh refinement, if the cell
contains gas mass larger than four times the DM particle mass scaled
by the baryon fraction.  In other words, the mesh is refined if the
cell contains the DM mass larger than $2(1-f_b)m_p$ or the gas mass
larger than $=4f_bm_p$ (where $m_p$ is given above and $f_b =
\Omega_{\rm b}/\Omega_{\rm m}$ = 0.1429).  We analyze clusters at the
present-day epoch as well as their progenitors at higher redshifts.

We repeated each cluster simulation with and without radiative
cooling.  The first set of ``adiabatic'' simulations have included
only the standard gasdynamics for the baryonic component without
dissipation and star formation.  The second set of simulations
included gasdynamics and several physical processes critical to
various aspects of galaxy formation: star formation, metal enrichment
and thermal feedback due to the supernovae type II and type Ia,
self-consistent advection of metals, metallicity dependent radiative
cooling and UV heating due to cosmological ionizing background
\citep{haardt_madau96}. We will use labels 'ad' and 'csf' for the
adiabatic simulations and simulations with cooling and star formation,
respectively. The cooling and heating rates take into
account Compton heating and cooling of plasma, UV heating, atomic and
molecular cooling and are tabulated for the temperature range
$10^2<T<10^9$~K and a grid of metallicities, and UV intensities using
the {\tt Cloudy} code \citep[ver. 96b4,][]{ferland_etal98}. The Cloudy
cooling and heating rates take into account metallicity of the gas,
which is calculated self-consistently in the simulation, so that the
local cooling rates depend on the local metallicity of the gas.

Star formation in these simulations was done using the
observationally-motivated recipe \citep[e.g.,][]{kennicutt98}:
$\dot{\rho}_{\ast}=\rho_{\rm gas}^{1.5}/t_{\ast}$, with
$t_{\ast}=4\times 10^9$~yrs. Stars are allowed to form in regions with
temperature $T<2\times10^4$~K and gas density $n > 0.1\ \rm cm^{-3}$.
No other criteria (like the collapse condition $\nabla\cdot {\bf v} <
0$) are used. We have compared runs where star formation was allowed
to proceed in regions different from our fiducial runs. We considered
thresholds for star formation of $n=10$, $1$, $0.1$, and $0.01\ {\rm
  cm^{-3}}$. We find that thresholds affects gas fractions at small
radii, $r/r_{\rm vir}<0.1$, but the differences are negligible at the radii we consider
in this study. The effect on the baryon fractions is small at all
radii.

Algorithmically, star formation events are assumed to
occur once every global time step $\Delta t_0\sim 10^7$ yrs, the value
close to the observed star formation timescales \citep[e.g.,][]{hartmann02}.
Collisionless stellar particles with mass
$m_{\ast}=\dot{\rho}_{\ast}\Delta t_0$ are formed in every unrefined
mesh cell that satisfies criteria for star formation during star
formation events. The mass of stellar particles is restricted to be
larger than $\min(5\times 10^7h^{-1}{\ \rm M_{\odot}},2/3\times m_{\rm
gas})$, where $m_{\rm gas}$ is gas mass in the star forming cell.
This is done in order to keep the number of stellar particles
computationally tractable, while avoiding sudden dramatic decrease of
the local gas density. In the simulations analyzed here, the masses of
stellar particles formed by this algorithm range from $\approx 10^5$
to $7\times 10^8h^{-1}{\ \rm M_{\odot}}$.

Once formed, each stellar particle is treated as a single-age stellar
population and its feedback on the surrounding gas is implemented
accordingly.  The feedback here is meant in a broad sense and includes
injection of energy and heavy elements (metals) via stellar winds and
supernovae and secular mass loss.  Specifically, in the simulations
analyzed here, we assumed that stellar initial mass function (IMF) is
described by the \citet{miller_scalo79} functional form with stellar
masses in the range $0.1-100\ \rm M_{\odot}$. All stars more massive
than $M_{\ast}>8{\ \rm M_{\odot}}$ deposit $2\times 10^{51}$~ergs of
thermal energy in their parent cell\footnote{No delay of cooling was
  introduced in these cells after SN energy release.} and fraction
$f_{\rm Z}= {\rm min}(0.2,0.01M_{\ast}-0.06)$ of their mass as metals,
which crudely approximates the results of \citet{woosley_weaver95}. In
addition, stellar particles return a fraction of their mass and metals
to the surrounding gas at a secular rate $\dot{m}_{\rm
  loss}=m_{\ast}\,\,C_0(t-t_{\rm birth} + T_0)^{-1}$ with $C_0=0.05$
and $T_0=5$~Myr \citep{jungwiert_etal01}. The code also accounts for
the SNIa feedback assuming a rate that slowly increases with time and
broadly peaks at the population age of 1~Gyr. We assume that a
fraction of $1.5\times 10^{-2}$ of mass in stars between 3 and $8\ \rm
M_{\odot}$ explodes as SNIa over the entire population history and
each SNIa dumps $2\times 10^{51}$ ergs of thermal energy and ejects
$1.3\ \rm M_{\odot}$ of metals into the parent cell. For the assumed
IMF, 75 SNII (instantly) and 11 SNIa (over several billion years) are
produced by a $10^4\ \rm M_{\odot}$ stellar particle.

Throughout this paper we use estimates of gas and baryon fractions
within different commonly used radii, defined by the total matter
overdensity they enclose. We will use radii $r_{2500}$, $r_{500}$,
$r_{200}$ enclosing overdensities of 2500, 500, and 200 with respect
to the critical density, $\rho_{\rm crit}$, as well as radii $r_{180}$
and $r_{\rm vir}$ enclosing overdensities of 180 and $\Delta_{\rm
  vir}$ with respect to the mean density of the universe. The latter
is equal to $\Delta_{\rm vir}\approx 334$ at $z=0$ and $\approx 200$
at $z=1$ for the cosmology adopted in our simulations.

The virial radius and masses of clusters within different radii are
given in Table~\ref{tab:sim}. For reference, we also give average
emission-weighted temperature in the radial range between $80$~kpc to
$r_{500}$ of each cluster in adiabatic and cooling runs. The
temperature is calculated as an emission-weighted average convolved
with the {\sl Chandra} energy response in the 0.5$-$7~keV energy band.

%-----------------------------------------------------
\section{Comparison of the ART and Gadget simulations}
\label{sec:artvsgadget}
%-----------------------------------------------------

\begin{figure}[tb]
 \vspace{-0.6cm}
 \hspace{0.2cm}
 \centerline{\epsfysize=4.75truein \epsffile{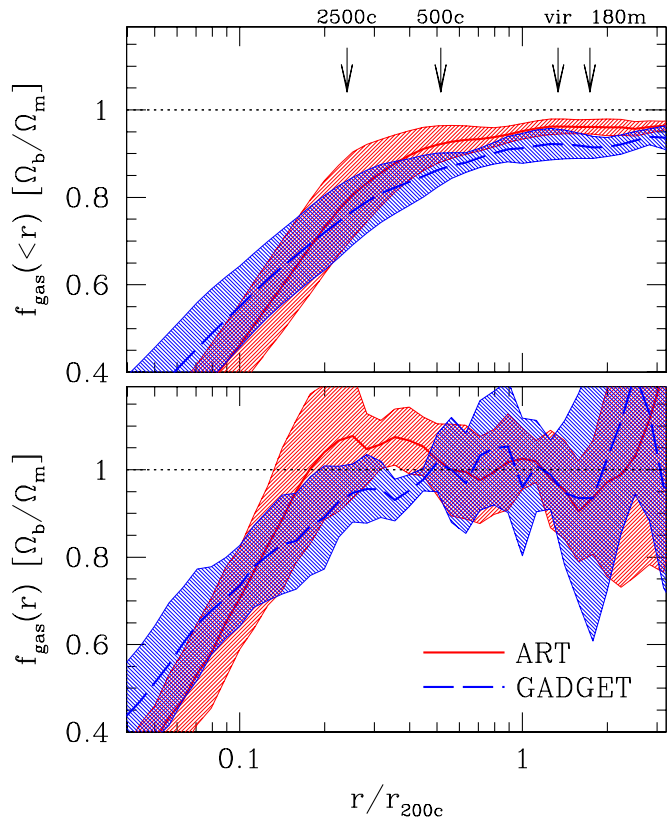}} 
 \vspace{-0.5cm}
\caption{Comparison of the baryon fraction profiles
  in the adiabatic simulations of the eight clusters CL2-9 done
  with the ART ({\it solid lines}) and entropy-conserving Gadget ({\it
    long-dashed lines}) codes. The gas fractions are normalized to the
  universal baryon fraction of each simulation, while radii of all
  clusters are normalized to their respective radius enclosing the
  overdensity of 200 with respect to the critical density. The {\it
    bottom panel} shows differential gas fraction profiles, while the
  {\it top panel} shows corresponding cumulative profiles. The shaded
  bands represent 1$\sigma$ {\sl rms} scatter around the mean for the
  eight clusters. The vertical arrows in the top panel show the radii
  enclosing overdensities of 2500, 500 (with respect to $\rho_{\rm
    crit}$), the virial overdensity and the overdensity of 180 with
  respect to the mean density. 
}
\label{fig:fg_artgadget}
\end{figure}

Before we begin, it is important to discuss the degree of numerical
uncertainty in the baryon fractions derived from cosmological
simulations. We will compare simulations of eight lower-mass clusters
(CL2-CL9) from the Table~\ref{tab:sim} performed with the eulerian AMR
Adaptive Refinement Tree code and the entropy conserving SPH Gadget
code \citep{springel_etal01,springel_hernquist02} performed in the
adiabatic regime.  The ART simulations are described above, while the
Gadget simulations are described by \citet{ascasibar_etal03}.  We
compare the CL2-CL9 for which the SPH simulations are available. In
each case the simulations have been started from the identical initial
conditions and ran in the same cosmology with comparable spatial and
mass resolutions, although Gadget simulations used somewhat different
values of $\Omega_{\rm b}$ for each individual cluster simulation. The
number of dark matter particles is the same in the ART and Gadget
simulations, while the number of gas particles varies but is of order
million for all clusters. Both gravitational and the minimum SPH
softening are in the range $\epsilon=2-5h^{-1}$~kpc, depending on
the number of particles within the virial radius.

The average cumulative and differential gas profiles in the two sets
of simulations are compared in Figure~\ref{fig:fg_artgadget}.  The
figure shows a rather remarkable agreement (to a couple percent) in
the differential profiles at $r\gtrsim 0.4r_{\rm vir}$, while at
smaller radii there are systematic differences. The gas fraction
profiles in the Gadget simulations are more centrally concentrated at
$r<0.2r_{\rm vir}$ compared to the ART simulations. At the same time,
the gas fraction is systematically higher in the ART simulations at
$r\sim 0.2-0.4r_{\rm vir}$. Comparison of the cumulative gas fraction
profiles shows that despite the fact that the ART profiles are less
centrally concentrated at small radii, the gas fraction within
$0.4r_{\rm vir}$ is higher than in the Gadget simulations by $\approx
5\%$.  At larger radii, $r\gtrsim r_{2500c}$, the systematic
difference persists, although it is only about $3-5\%$. Note also that
the scatter about the mean is similar in the ART and Gadget
simulations.  We checked that the differences in the mean profiles of
the eight clusters also exist between profiles of individual
clusters. They are therefore systematic and statistically significant.

The difference is significantly smaller than the systematic difference
in gas fractions of $\approx 10\%$ between the eulerian and SPH codes,
observed in the Santa Barbara cluster comparison project
\citep{frenk_etal99}. Indeed, agreement in gas and dark matter density
profiles at this level is quite remarkable. Nevertheless, it is
important to keep these systematic differences in mind, both in
gauging implications of the results presented below and in
observational analyses of gas fractions for cosmological constraints.
As we argue below, this systematic difference is smaller than the
uncertainty in the simulations with cooling and star formation,
associated with the uncertain fraction of baryons in stars.

%------------------------------
\section{Effects of Resolution}
\label{sec:resolution}
%------------------------------

Given that in these simulations we are trying to resolve galaxy
formation on kiloparsec scales while following formation of clusters
self-consistently on scales of megaparsecs, it is reasonable to ask
how the finite dynamic range of the simulations affects our results.
To this end, we have re-run one of the clusters used in our study
(CL3) with higher resolution.  The re-simulations use more agressive
refinement criteria, which results in a factor of $\simeq 2.5$ more
refinement cells and a factor of two higher spatial resolution
compared to the run used in this study. For example, the
higher-resolution simulations at $z=0$ has $\approx 2\times 10^6$ grid
cells on the 9th refinement level (comoving cell size of
$1.22h^{-1}$~kpc), while the runs used in this study did not have
refinement cells beyond the 8th level. Gas fraction profile in the
adiabatic re-simulation with such higher resolution agrees with the
lower resolution simulation to within $\simeq 1\%$ at all resolved
radii.  We find a similarly small difference for the higher-resolution
re-simulation of this cluster with the Gadget code with eight times
more particles and softening of $0.5h^{-1}$~kpc (compared to the
$2-5h^{-1}$~kpc in the simulations we use for comparison). In other
words, it appears that for each code convergence in adiabatic runs is
reached, so that different resolution does not explain systematic
differences in the gas fraction profiles discussed above.

The higher-resolution re-simulation of CL3 with cooling and star
formation results in total baryon fraction profile that is within
$\simeq 1-2\%$ of that in the low-resolution simulation.  However, we find
that the gas fraction is $\simeq 3-5\%$ higher in the higher resolution
run. It is not clear what causes the difference, although we think
that it is most likely be due to the increased heating by stellar
feedback in the higher-resolution simulation. The difference is 
relatively small and is considerably smaller than the magnitude of the
effects we discuss. Nevertheless, the effects of resolution
in the runs with star formation will have to be studied further 
with higher-resolution runs.

%-------------------
\section{Results}
\label{sec:results}
%-------------------

We will start discussion of our results by considering an example of
a single typical cluster from our sample (CL3).  The cluster has virial mass
similar to that of the Virgo cluster and has experienced a nearly
equal-mass major merger at $z\approx 0.6$.  We have analyzed three
simulations of this cluster started from the same initial conditions
but run including different physical processes. The first run was done
in the ``adiabatic'' regime without radiative gas cooling and star
formation.  The second run included cooling and star formation.  The
third run is identical to the second at $z>2$. At $z<2$, however, gas
dissipation and star formation have been artificially turned off.
This run should be considered not as a realistic cluster model, but as
a useful intermediate case between adiabatic run and run with full
dissipation. For
example, the fraction of baryons turned into stars within the virial
radius is $f_{\ast}=0.35$ in the simulation with
cooling and $f_{\ast}=0.15$ in the
simulation with no cooling at $z<2$ ($f_{\ast}=0$ for the adiabatic
simulation).

\begin{figure}[tb]
 \vspace{-0.6cm}
 \hspace{0.2cm}
 \centerline{\epsfysize=4.75truein \epsffile{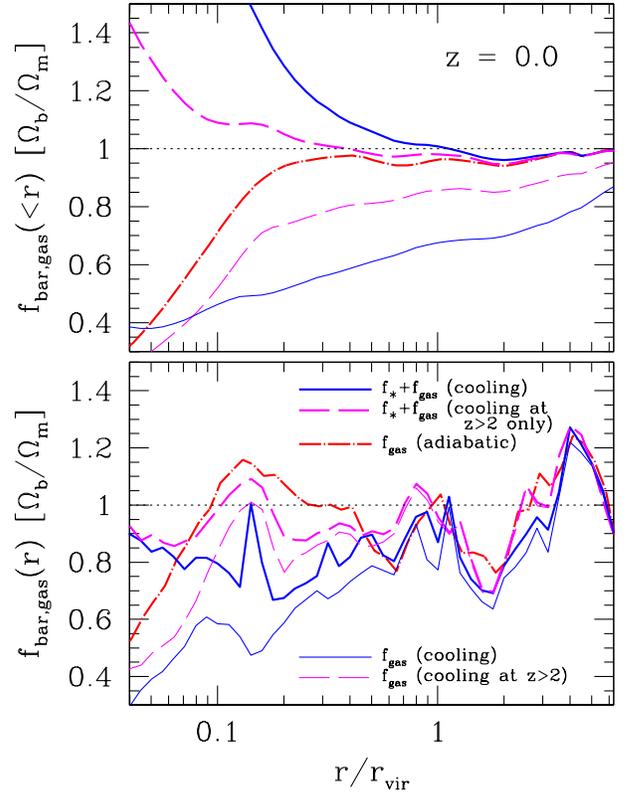}} 
 \vspace{-1.0cm}
\caption{Effects of cooling on the cluster baryon and gas fraction profiles
  at $z=0$. The figure shows profiles for the three resimulations of
  the Virgo-size cluster CL3. {\it Top panel:} the cumulative baryon
  and gas fractions within a given radius in the simulations with
  cooling ({\it solid}, {\it thick lines} show $f_{\rm
    gas}+f_{\ast}$ while {\it thin lines} show $f_{\rm gas}$), cooling
  only at $z>2$ ({\it long-dashed lines}), and no cooling (or
  adiabatic, {\it dot-dashed line}).  The radius is in units of the
  virial radius, defined as the radius enclosing overdensity of
  $\Delta_{\rm vir}=334$ with respect to the mean density of the
  universe. The fractions are in units of the universal baryon
  fraction, $\Omega_b/\Omega_m$.  {\it Bottom panel:} the same as the
  top panel, but for the differential profiles. }
\label{fig:fbg_cl6}
\end{figure}

Figure~\ref{fig:fbg_cl6} shows the cumulative and differential
profiles of baryon and gas fraction for the three re-simulations of
CL3. The figure shows several key qualitative effects of dissipation.
The most salient difference is in the gas fraction. The simulation
with cooling has the largest stellar fraction and, correspondingly,
the smallest $f_{\rm gas}(<r_{\rm vir})$. The difference in cumulative
gas fraction is approximately constant at $r\gtrsim 0.15r_{\rm vir}$.
However, the lower panel of Figure~\ref{fig:fbg_cl6} shows that the
difference in differential profile of $f_{\rm gas}$ increases
monotonically with decreasing radius from less than 10\% at $r\gtrsim
0.4r_{\rm vir}$ to $\sim 30-40\%$ at smaller radii. This is because
the central cluster galaxy has a significant effect on the
overall stellar fraction and cumulative profile. 

Comparison of the total baryon fraction profiles for the three runs
reveals additional differences. The baryon fraction in the adiabatic
simulation is significantly smaller than the universal value at
$r\lesssim 0.2r_{\rm vir}$, while it is larger than universal in the
simulations with cooling. The cumulative baryon fraction in
the adiabatic simulation reaches the universal value well beyond the
virial radius. In the simulations with cooling the baryon fraction is
close to the universal at the virial radius.  Note that in all three
runs the baryon fractions are within $10\%$ of the universal value at
$r\gtrsim 0.4r_{\rm vir}$.  The baryon fraction profiles are similar
in all three simulations outside the virial radius.

The main conclusion from this comparison, which also applies to the
comparison of the mean profiles presented below, is that the cooling
and star formation change not only the amplitude but also the shape of
the gas and baryon fraction profiles in the simulated clusters.  

This is to be expected because cooling and star formation lead to
redistribution of both baryons and dark matter. The gas condensation
results in a contraction of the dark matter halo
\citep[e.g.,][]{zeldovich_etal80,blumenthal_etal86} and steepening of
the DM density profile at $r\lesssim 0.1r_{\rm vir}$
\citep{gnedin_etal04}. More significantly, dissipation and resulting
star formation convert a fraction of gas into stars with the
efficiency which may depend on the cluster radius. Dynamically, the
stellar component is collisionless and its evolution will differ from
that of the gas. For instance, cluster galaxies may loose their gas
via ram pressure stripping. The lost gas can then mix with the
intracluster gas at large radii. The stellar components of galaxies,
on the other hand, can sink to the center via dynamical friction and
build up the massive central cluster galaxy. Likewise, the detailed
dynamics and relaxation process of gas and stars during mergers is
rather different.  For example, the gas experiences heating via
shocks, while stars participate in collisionless violent relaxation.

\begin{figure*}[t]
 \vspace{-0.8cm}
 \hspace{-0.4cm}
 \centerline{\epsfysize=6.2truein \epsffile{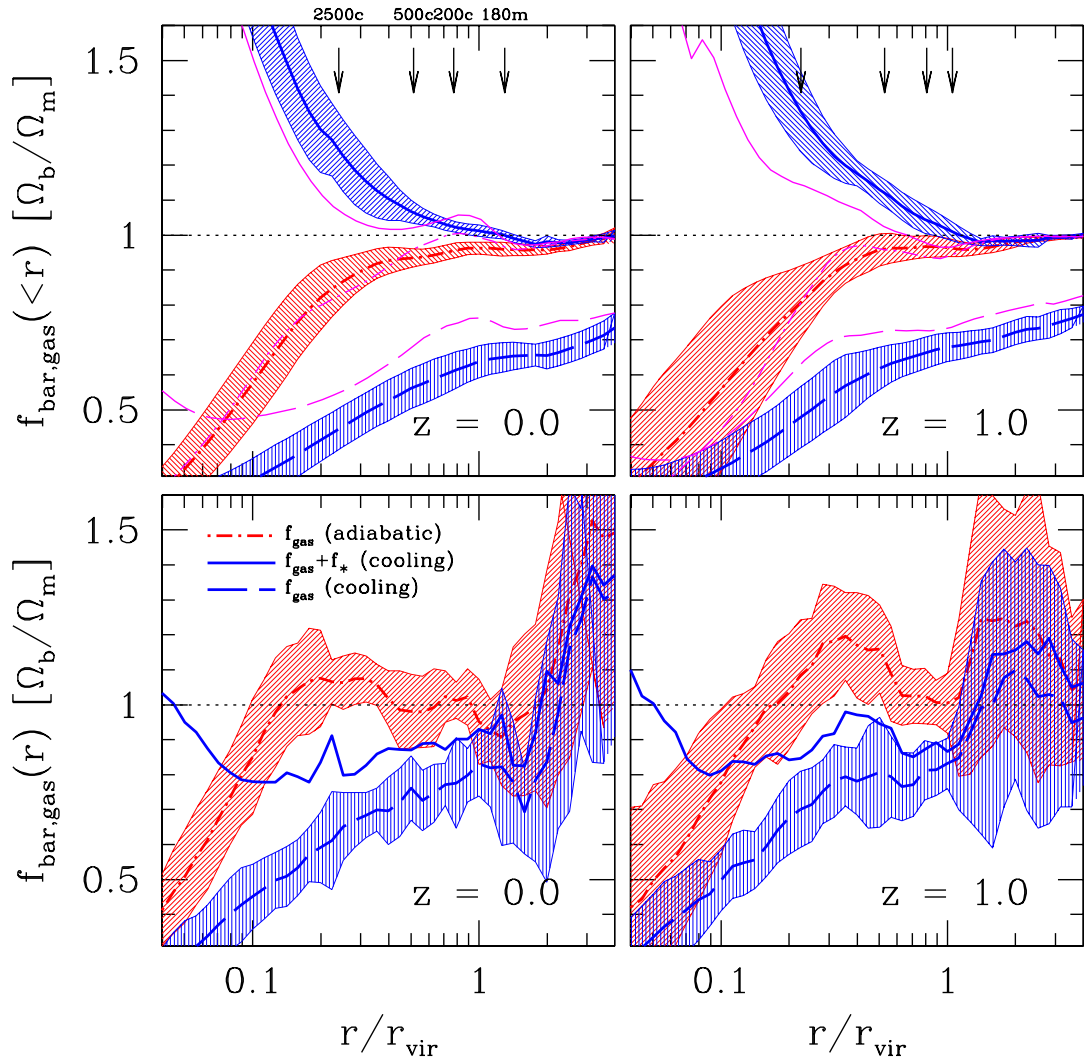}} 
 \vspace{-.5cm}
\caption{Differential ({\it bottom panels}) and cumulative ({\it top panels})
  baryon and gas fraction profiles for the nine clusters used in our
  analysis at $z=0$ ({\it left column}) and $z=1$ ({\it right
    column}). The {\it dot-dashed} lines show the mean profiles in the
  adiabatic simulations averaged over eight clusters (CL2-CL9).  The
  {\it solid} lines show the total mean baryon fraction (gas$+$stars)
  profile, while the {\it long-dashed} lines show the gas fraction
  profile in the simulations with cooling and star formation. The
  shaded bands show the 1$\sigma$ {\sl rms} scatter around the mean
  for the eight clusters. The {\it thin lines} in the top panels show
  the corresponding profiles in the Coma-size cluster simulation
  (CL1). These profiles have not been used in the averages due to
  systematically different gas and baryon fractions. The vertical
  arrows in the top panels show the radii enclosing overdensities of
  2500, 500, 200 (with respect to $\rho_{\rm crit}$), and the
  overdensity of 180 with respect to the mean density. }
\label{fig:fbg}
\end{figure*}

\begin{figure*}[tb]
 \vspace{-2cm}
 \hspace{-0.4cm}
 \centerline{\epsfysize=7.2truein \epsffile{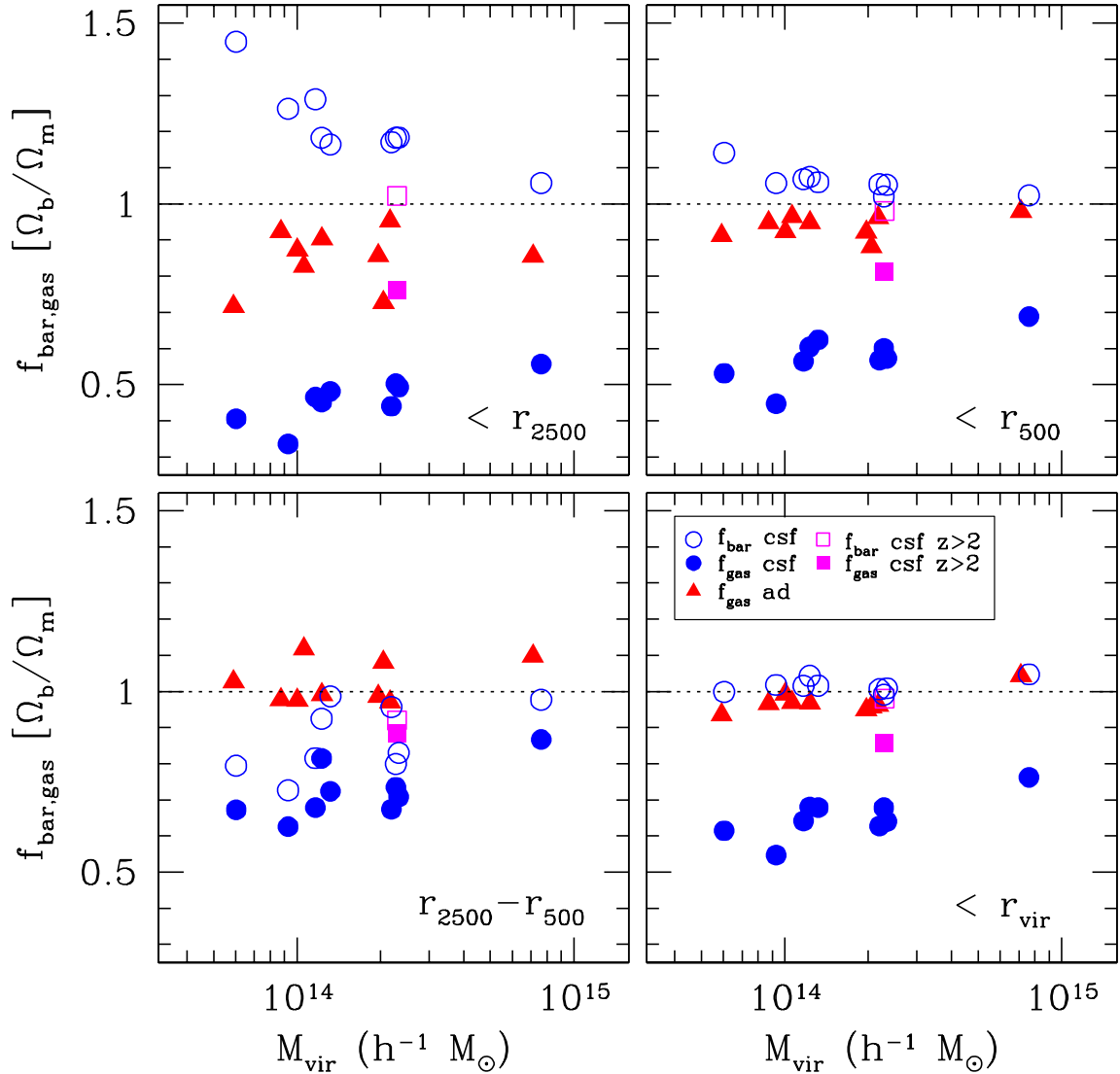}} 
 \vspace{-1.0cm}
\caption{The total baryon and gas fractions of individual clusters in
  units of the universal value within different radii (clockwise from
  the top left panel: $r<r_{2500}$, $r<r_{500}$, $r<r_{\rm vir}$, in
  the annulus $r_{2500}-r_{500}$) as a function of the cluster virial
  mass. {\it Solid triangles} show the gas fractions in the adiabatic
  simulations. The {\it solid circles} show the gas fraction, while
  {\it open circles} show the total baryon fraction in the simulations
  with cooling and star formation. {\it Solid} and {\it open squares}
  show the gas and baryon fractions in the resimulation of CL3, in
  which cooling was turned off at $z<2$. Note the systematic trend
  with mass for the gas fractions in the simulation with cooling. }
\label{fig:fbgmv}
\end{figure*}

\begin{figure*}[t]
 \vspace{-0.2cm}
 \hspace{-0.4cm}
 \centerline{\epsfysize=6.2truein \epsffile{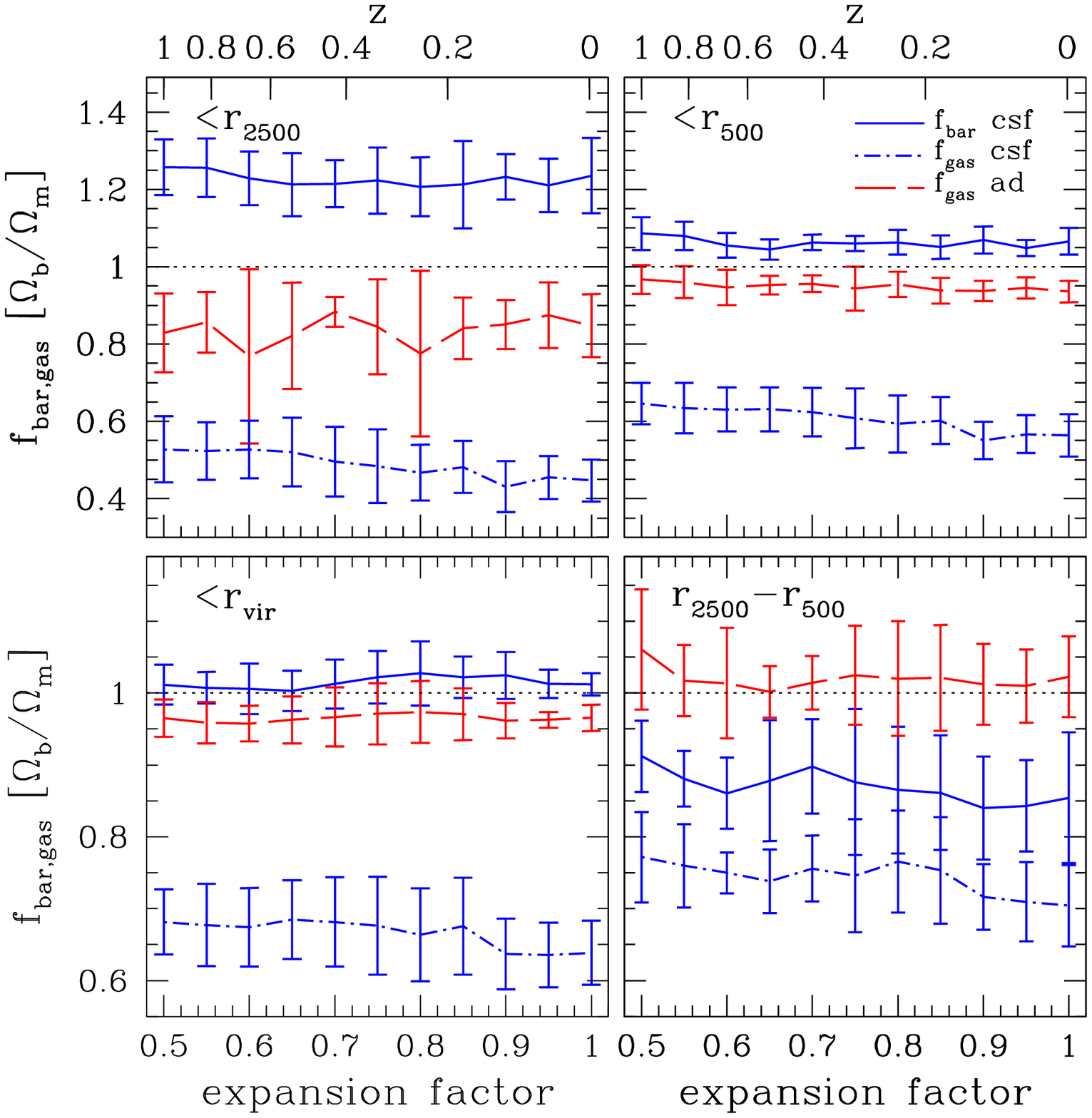}} 
 \vspace{-0.5cm}
\caption{Evolution of the baryon and gas fractions within different
radii for the eight clusters CL2-CL9 from $z=1$ to $z=0$. The 
radii used are the same as in Fig.~\ref{fig:fbgmv}. The {\it long-dashed}
lines shows the gas fraction in the adiabatic simulations. The {\it dot-dashed} and {\it solid} lines show the gas and baryon fraction profiles 
in the simulation with cooling and star formation. The gas and baryon
fractions are normalized to the universal fraction of baryons assumed
in the simulations. The figure shows that the evolution of baryon fraction
at these redshifts is weak. Note that each cluster at least doubles
its mass over this redshift interval. }
\label{fig:fbga}
\end{figure*}

Figure~\ref{fig:fbg} shows the comparison of the mean baryon and gas
fraction profiles at $z=1$ and $z=0$ for our entire sample of clusters
simulated in the adiabatic regime and with dissipation. The profile
for the Coma-size cluster (CL1) is plotted separately because in this
cluster stellar fraction (and hence the gas fraction) is significantly
different from the rest of the smaller-mass clusters, as we discuss
below. However, the difference is mainly in the amplitude of the
profile. Qualitatively, the profiles of CL1 are similar to the mean profile
for the other eight clusters.

The trends in the mean profiles are similar to those seen in
Figure~\ref{fig:fbg_cl6}.  The cumulative baryon fraction profiles are
flat at $r\gtrsim 0.4r_{\rm vir}$. At these radii $f_{\rm bar}$ is
within $\approx 5\%$ of the universal value for both adiabatic and CSF
runs, although in the former the $f_{\rm bar}$ is systematically below
the universal value, while in the latter it is systematically above.
In the cores of the low-mass clusters, the cumulative baryon
fraction is considerably below (above) the universal value in the
adiabatic (CSF) runs. This shows that dissipation and star formation
result in a substantial redistribution of baryons within the
virial radius. In the Coma-size cluster the baryon fraction is still
close to the universal value within $r_{2500}$, but at smaller radii
the trends are similar.  Note that the differential baryon fraction in
the CSF simulations is systematically below the universal value by
$\approx 10-20\%$. The profile, however, is rather flat at $r\gtrsim
r_{2500}$.  Both the cumulative and differential gas fractions
increase monotonically with radius. If a similar trend exists in real
clusters this implies that the correction of gas fraction for stars in
general depends on the cluster-centric radius and a single assumed
value of $f_{\ast}$ for all radii may not give the universal baryon
fraction.

Figure~\ref{fig:fbg} also shows that the variance around the mean for
the cumulative profiles is rather small, especially at large radii.
The scatter is somewhat larger at $z=1$, probably because the clusters
are on average less relaxed at the earlier time. However, there is no
significant qualitative difference between the profiles at two
different epochs.

Figure~\ref{fig:fbgmv} shows the baryon and gas fractions as a
function of the cluster virial mass within
$r_{2500}$, $r_{500}$, and $r_{\rm vir}$ and 
within the radial range $[r_{2500}-r_{500}]$. The values of the baryon
and gas fractions plotted in this figure are listed in
Table~\ref{tab:fbg}. The fractions in the adiabatic simulations are
marked 'ad', while those in the simulations with cooling and star
formation are marked 'csf' (e.g., $f^{\rm csf}_{\rm gas}$). In the
table we also provide the mean values and scatter for our cluster
sample. Note, however, that these are averages over clusters of
different masses and should be interpreted with caution given the
trends of the baryon and gas fractions with cluster mass discussed
below.

In adiabatic simulations, the baryon fraction appears to be
independent of mass for all radial ranges. It is systematically below
the universal value by $\approx 15\%$ and $\approx 5\%$ within
$r_{2500}$ and $r_{500}$, respectively.  For $r<r_{\rm vir}$, the
baryon fraction is within $\approx 5\%$ of the universal value.  Note
also that in the adiabatic runs, the baryon fraction within the radial
range of $r_{2500}-r_{500}$ is close to the universal value on
average, even though the cumulative fractions within these radii are
systematically below.

In the simulations with cooling and star formation the baryon and
gas fractions show (weak) trend with mass. The gas fraction appears
to systematically increase with increasing cluster mass as 
\begin{equation}
f_{\rm gas}=0.2\log(M_{\rm vir}/10^{14}h^{-1}{\rm\ M_{\odot}}) + f_0,
\label{eq:fgas_fit}
\end{equation}
where $f_0$ is $\approx 0.4$, $0.5$, $0.6$, $0.65$ for $r<r_{2500}$,
$r_{500}$, $r_{\rm vir}$, and $r_{2500}-r_{500}$, respectively. The
significance of this trend and the value of the slope are uncertain
given the small size of the sample. The baryon fraction in the CSF
runs is larger than the universal value within $r_{500}$, but it is
within $\approx 2-3\%$ of the universal value within $r_{\rm vir}$.
The baryon fraction, $f_{\rm bar}(<r_{\rm vir})$, in these runs is
also not biased low like the corresponding fraction in the adiabatic
runs but gives the average very close to the universal.

The differential fractions within the radial range $r_{2500}-r_{500}$
behave somewhat differently because they are not affected by the
concentration of baryons in the cluster center. In this radial range,
the baryon fraction in the adiabatic runs is approximately unbiased.
In the CSF runs the baryon fraction is systematically below the
universal value but its mass dependence for $M\gtrsim
10^{14}h^{-1}{\rm\ M_{\odot}}$ seems to be weaker than for the
cumulative fraction within similar radii. In all panels open and solid
squares show the baryon and gas fractions for the re-simulation of CL3
with cooling turned off at $z<2$.  As could be expected, the
fractions in this case are intermediate between the adiabatic
simulation and run with cooling and star formation.

Finally, in Figure~\ref{fig:fbga} we show evolution of the average
baryon and gas fractions from $z=1$ to $z=0$ within radii shown in the
previous figure in the adiabatic and CSF runs of clusters CL2-CL9.
The figure shows that the total baryon fractions are quite stable in
this redshift interval. This is remarkable given that most
clusters undergo at least one major merger at $z<1$.  There is a weak
decrease of the gas fraction with time in the CSF simulations, as some
of the gas is cooling and is turned into stars. The evolution of
$f_{\rm gas}$ is $\approx 10-20\%$ from $z=1$ to $z=0$. 

%------------------------------------
\section{Comparison to previous work}
\label{sec:previous}
%-----------------------------------

Given the systematic differences in gas fractions in the Santa Barbara
cluster comparison \citep{frenk_etal99}, it is interesting to compare
the results presented in the previous section to other recent studies
to assess the current state of affairs.  The comparison is relatively
straightforward for the adiabatic simulations because all codes should
simulate evolution using the same equations of gas and collisionless
dynamics. Any differences can therefore be attributed to the
difference in the actual numerical scheme. \citet{eke_etal98} studied
SPH simulations of ten massive clusters formed in the concordance
$\Lambda$CDM cosmology.  They found gas fractions of $\approx 0.99$,
$0.87$, and $0.83$ within $3r_{\rm vir}$, $r_{\rm vir}$, and
$0.5r_{\rm vir}$ (note that $r_{500}\approx 0.5r_{\rm vir}$),
respectively, which evolve only very weakly from $z=1$ to $z=0$.
These numbers are similar to those obtained with a number of other SPH
codes \citep{frenk_etal99,bialek_etal01,muanwong_etal02}.  We also
find no detectable evolution of baryon fraction in adiabatic
simulations over this period of time.  However, the gas fractions in
our simulations are larger: $\approx 1.00$ at $r\gtrsim 3r_{\rm vir}$,
and $0.97\pm 0.03$ and $0.94\pm 0.03$ for $r<r_{\rm vir}$ and
$r<r_{500}$. Gas fractions in simulations with the entropy-conserving
Gadget code (see Fig.~\ref{fig:fg_artgadget}) are larger by about 5\%
than those found by \citet{eke_etal98}: $\approx 0.92$ at $r<r_{\rm
  vir}$ and $\approx 0.88$ if measured within $0.5r_{\rm vir}$.
This is consistent with the adiabatic simulations (run with the 
entropy conserving version of the Gadget) presented in 
a recent study by \citet{kay_etal04}, who find the average 
gas fraction of $\simeq 90\%$ of the universal value within 
the virial radius. The average gas fraction profile for the adiabatic 
simulations in their Figure~6 is in good agreement with the Gadget
simulations presented here. 

\begin{deluxetable*}{ccccccccccccc}[t]
\tablecaption{Baryon and gas fractions in cluster simulations at the present day epoch}
\tablehead{\\
\multicolumn{1}{c}{}&
\multicolumn{3}{c}{$r<r_{2500}$}&
\multicolumn{3}{c}{$r<r_{500}$}  &
\multicolumn{3}{c}{$r_{2500}-r_{500}$}  &
\multicolumn{3}{c}{$r<r_{\rm vir}$}  
\\
\colhead{Cluster}&
\colhead{$f^{\rm ad}_{\rm bar}$} &
\colhead{$f^{\rm csf}_{\rm gas}$}&
\colhead{$f^{\rm csf}_{\rm bar}$} &
\colhead{$f^{\rm ad}_{\rm bar}$} &
\colhead{$f^{\rm csf}_{\rm gas}$}&
\colhead{$f^{\rm csf}_{\rm bar}$} &
\colhead{$f^{\rm ad}_{\rm bar}$} &
\colhead{$f^{\rm csf}_{\rm gas}$}&
\colhead{$f^{\rm csf}_{\rm bar}$} &
\colhead{$f^{\rm ad}_{\rm bar}$} &
\colhead{$f^{\rm csf}_{\rm gas}$}&
\colhead{$f^{\rm csf}_{\rm bar}$} \\
}
\startdata
\\
CL1    & 0.86 & 0.56 & 1.06 & 0.98 & 0.69 & 1.02 & 1.10 & 0.87 & 0.98 & 1.04 & 0.76 & 1.05 \\
CL2    & 0.73 & 0.44 & 1.17 & 0.88 & 0.57 & 1.05 & 1.08 & 0.67 & 0.79 & 0.96 & 0.63 & 1.01 \\
CL3    & 0.95 & 0.49 & 1.18 & 0.96 & 0.57 & 1.05 & 0.97 & 0.71 & 0.83 & 0.96 & 0.64 & 1.01 \\
CL4    & 0.86 & 0.50 & 1.18 & 0.92 & 0.60 & 1.02 & 0.99 & 0.74 & 0.80 & 0.95 & 0.68 & 0.99 \\
CL5    & 0.90 & 0.48 & 1.16 & 0.95 & 0.62 & 1.06 & 0.99 & 0.72 & 0.99 & 0.97 & 0.68 & 1.02 \\
CL6    & 0.92 & 0.34 & 1.26 & 0.95 & 0.45 & 1.06 & 0.98 & 0.63 & 0.73 & 0.97 & 0.55 & 1.02 \\
CL7    & 0.87 & 0.45 & 1.18 & 0.92 & 0.60 & 1.08 & 0.98 & 0.82 & 0.92 & 0.99 & 0.68 & 1.04 \\
CL8    & 0.83 & 0.47 & 1.29 & 0.97 & 0.56 & 1.07 & 1.12 & 0.68 & 0.82 & 0.97 & 0.64 & 1.02 \\
CL9    & 0.72 & 0.41 & 1.45 & 0.91 & 0.53 & 1.14 & 1.03 & 0.67 & 0.79 & 0.94 & 0.61 & 1.00 \\
\\
mean   & 0.85 & 0.46 & 1.22 & 0.94 & 0.58 & 1.06 & 1.02 & 0.72 & 0.87 & 0.97 & 0.65 & 1.02 \\
scatter& 0.08 & 0.06 & 0.11 & 0.03 & 0.07 & 0.03 & 0.06 & 0.08 & 0.09 & 0.03 & 0.06 & 0.02 \\
\enddata
\label{tab:fbg}
\end{deluxetable*}

The studies of baryon fraction in simulations with cooling and star
formation are fewer and the source of the differences is usually not
as clear, because it can be attributed both to a different
implementation of these processes and to the difference between
numerical gasdynamics schemes.  \citet[][]{muanwong_etal02} compared
the baryon and gas fractions as a function of cluster virial mass in
adiabatic simulations and re-simulations in which gas was allowed to
cool or was pre-heated (see their Fig. 3). For the mass range of our
cluster sample, in simulations with cooling \citet{muanwong_etal02}
obtain baryon fractions of $\approx 0.85-0.9$ of the universal value
within the virial radius. This is slightly larger than in their
adiabatic simulations, but is considerably lower than in our
simulations with cooling and star formation ($1.02\pm 0.02$, see
Table~\ref{tab:fbg}). \citet{valdarnini03} compares gas fractions in
in TREESPH simulations that include cooling and star formation to the
observational estimates of \citet{arnaud_evrard99} and finds a
reasonable agreement. \citet{valdarnini03} also finds indications of a
trend of increasing $f_{\rm gas}$ with increasing cluster mass similar
to the trend we find in our simulations. The actual gas fractions in
this study are, however, somewhat lower than our values, which may be
due to a lower assumed universal baryon fractions ($\Omega_{\rm
  b}h^2=0.015-0.019$) compared to the value assumed in this study.
\citet{kay_etal04} analyze a sample of 15 clusters simulated using the
entropy-conserving Gadget code with cooling, star formation, and
feedback. Their average gas fraction profile for these simulations
agree well with the gas fraction profile of our small-mass clusters
(CL2-CL9), but is considerably lower than that for our Coma-size
cluster (CL1). Most recently, \citet{ettori_etal04} presented a
similar analysis for the simulations with cooling and star formation
of a statistical sample of clusters simulated with the
entropy-conserving version of the Gadget code. These authors,
quote total baryon fractions of $f_{\rm bar}(<r_{\rm vir})\approx
0.93-0.95$ of the universal value in their simulations. This is closer
to the values in our simulations. Their gas fractions are $0.75-0.8$
--- higher than in our clusters due to a different implementation of
cooling and star formation.

These comparisons and the comparison with the Gadget code presented in
\S~\ref{sec:artvsgadget} indicate that there are still systematic
differences between the eulerian and SPH codes, with the latter
predicting systematically smaller baryon fractions. The difference
appears to be the smallest when eulerian simulations are compared to
the entropy-conserving Gadget SPH code.

%-----------------------------------
\section{Discussion and conclusions}
\label{sec:discussion}
%-----------------------------------

We have presented results of the analysis of baryon and gas fractions
in the high-resolution simulations of nine galaxy clusters formed in
the concordance $\Lambda$CDM cosmology.  The clusters span the range
of virial masses from $M_{\rm vir}=6\times 10^{13}$ to $8\times
10^{14}h^{-1}{\rm\ M_{\odot}}$ and have {\it not} been selected to be
in the highly relaxed state at the present epoch. We study the effects
of cooling and star formation by comparing simulations of individual
clusters done with and without these processes included.

We show that radiative cooling of gas and subsequent star formation
significantly modify distribution of baryons within the virial radii
of galaxy clusters in our simulations. The effect is twofold. First,
cooling results in a condensation of gas and dark matter in the
centers of halos
\citep{zeldovich_etal80,barnes_white84,blumenthal_etal86,jesseit_etal02,gnedin_etal04}.
As the gas cools in the cluster progenitor, the baryons condense near
the cluster center (the central cluster galaxy). The condensing gas
has to be replaced by the gas from larger radii.  The net result is
that the baryon fraction is affected at radii up to $\sim r_{\rm
  vir}$. For example, Figure~\ref{fig:fbg} shows that the cumulative
baryon fraction in the simulations with cooling is systematically
higher than in the adiabatic runs at $r\lesssim 1.5r_{\rm vir}$ with
the difference of $\approx 5\%$ at $r_{\rm vir}$. Dissipation thus
allows clusters to collect a larger fraction of baryons within their
virialized region.  Remarkably, in these simulations baryon fraction
within the virial radius is very close (to $\approx 2-3\%$) to the
universal value.  Note that for the differential baryon fraction,
$f_{\rm bar}(r)$, the effect at large radii is opposite. In the
simulations with cooling $f_{\rm bar}(r)$ is systematically smaller
than in the adiabatic simulations at $r>0.1r_{\rm vir}$. The overall
increase in the cumulative baryon fraction is therefore driven by the
baryons in the central cluster galaxy.

Second, a fraction of gas is converted into collisionless stellar
component.  The ratio of stellar to gas mass depends on the distance
to the cluster center.  In cluster cores ($r\lesssim 0.2r_{\rm vir}$)
stellar fraction dominates the baryon budget, $f_{\ast}\gtrsim
50-70\%$ depending on cluster mass. At larger radii most baryons
remain in the form of gas with $f_{\ast}\lesssim 10-20\%$ at
$r>r_{2500}$.  The collisionless dynamics of stars is different from
the dynamics of gas in detail. For example, the gas in a galaxy or a
group can be ram-pressure stripped and deposited at larger radii,
while the stars can sink to the center via dynamical friction. During
mergers, the gas is shock-heated, while stellar particles relax via
the violent relaxation. When stars are formed, a smaller fraction of
baryons is subject to strong shocks that may prevent accretion of some
of the gas into the inner regions of clusters.  The differences in
dynamics can modify the total baryon fraction profile and the ratio of
stellar to gas mass at different radii.

To gauge how robust are the predictions of numerical simulations, we
have compared the gas fraction profiles for the eight clusters in our
sample, simulated in the adiabatic regime with the entropy-conserving
Gadget and the ART codes. The profiles in the two sets of simulations
agree to better than $\approx 3\%$ outside the cluster core ($r/r_{\rm
  vir}\gtrsim 0.2$), but differ by up to 10\% at small radii.  The
main difference is thus in the cluster core where the gas profiles in
the Gadget simulations are more centrally concentrated and the gas
fractions are correspondingly larger than in the ART runs.  This
is reflected in systematic difference in the
cumulative baryon fraction profiles at $r\gtrsim r_{2500}$: 95\% in
the ART runs compared to 92\% in the Gadget simulations. The
discrepancy is smaller than the difference in gas fractions of $\approx
10\%$ between SPH and eulerian codes in the Santa Barbara comparison
project \citep[][]{frenk_etal99}.  Nevertheless, it is systematic
(similar for all eight compared clusters) and one has to keep it in
mind when interpreting or using results of numerical simulations. Note
also that this comparison was done for the simplest case of the
adiabatic gas dynamics. The differences in baryon and gas fractions
are likely to be larger when cooling and star formation are included.
We discussed how our results compare with some of the recent
simulation results in the previous section.  We also plan to compare
our results with the results of the Gadget simulations with cooling
and star formation in the near future.

One should note that cluster simulations may suffer from the
``overcooling problem.''  Indeed, the fraction of baryons in the cold
gas and stars within the virial radius of clusters at $z=0$ in our
simulations is in the range $\sim 0.25-0.35$, at least a factor of two
higher than observational measurements for the systems of the mass
range we consider \citep{lin_etal04}. Such high fractions of baryons
in condensed cold gas or stars are generic results of cosmological
simulations, although specific numerical values vary
\citep[e.g.,][]{suginohara_ostriker98,lewis_etal00,pearce_etal00,dave_etal02,ettori_etal04}.
We have done an extensive convergence study, varying the numerical
resolution, implementation of cooling, star formation and stellar
feedback, and assumed baryon fraction. However, we have not been able
to reduce stellar fractions considerably below the values presented in
this study.
 
Note that \citet{motl_etal04} have recently presented eulerian AMR
simulations of two massive galaxy clusters ($M_{\rm vir}\approx
2\times 10^{15}h^{-1}{\rm\ M_{\odot}}$) with the fraction of cold
($T<15000$~K) condensed gas of $\approx 10\%$. It is not yet clear
whether this represents a discrepancy with our results. The number of
simulated massive clusters is clearly small.  At the same time, the two
simulated clusters analyzed by \citet{motl_etal04} have mass a factor
of two larger than our most massive cluster. If the trend of increasing
$f_{\rm gas}$ (and correspondingly decreasing $f_{\ast}$) with
increasing cluster mass (see Fig.~\ref{fig:fbgmv}) continues to larger
masses, equation~(\ref{eq:fgas_fit}) gives $f_{\rm gas}\approx
90\%$ for the cluster masses considered by \citet{motl_etal04},
in agreement with their results.

The efficient dissipation and persistence of late star formation may
indicate that some mechanism suppressing gas cooling is needed. The
problem appears to be in the central brightest cluster galaxy, which
contains a larger fraction of cluster mass than is measured in
observations.  The rest of the cluster galaxies have stellar masses in
reasonable agreement with observations (Nagai \& Kravtsov, in
preparation).  The overcooling problem in simulations may thus be
related to the puzzling absence of cold gas in the observed cluster
cores \citep[e.g.,][]{peterson_etal03}. At the same time, some
fraction of cluster stellar mass in low-surface brightness
intracluster component may be missed in observations, which could
reduce the apparent discrepancy of observed $f_{\ast}$ with the
simulation results
\citep[e.g.,][]{gonzalez_etal04,lin_mohr04}. For example, baryon
fractions within virial radii of rich clusters do not seem to add up
to the universal value from the WMAP observations
\citep{spergel_etal03}, as they do in both adiabatic and dissipative
simulations \citep[see, e.g.,][]{ettori03}.  

Interestingly, gas fractions measured in our CSF simulations are
consistent with the observed gas fractions in clusters
\citep{evrard97,mohr_etal99,arnaud_evrard99,roussel_etal00}. For example,
\citet{evrard97} presents estimates of the gas fractions within radius
of $r_{500}$ for a sample of galaxy clusters spanning a wide range of
the ICM temperatures. The gas fractions vary from $\sim 50$ to $85\%$
of the universal value (assuming $\Omega_{\rm b}=0.043$ and $h=0.7$)
in rough agreement with our results.  The values of gas fraction in
units of the universal value, $f_{\rm bar}=0.143$, estimated from the
virial scaling relations by \citet{mohr_etal99} are\footnote{We have
calculated the gas fractions for $h=0.7$.}  $f_{\rm
gas}(<r_{500})/f_{\rm bar}\sim 0.5-0.7$ for clusters with temperatures
of $T_X\approx 4$~keV and $f_{\rm gas}(<r_{500})/f_{\rm bar}\sim
0.65-0.85$ for $T_X\approx 8$~keV.  The latter is also consistent with
the measurements based on the SZ cluster observations and X-ray
temperature of $T_X\gtrsim 5$~keV clusters \citep{grego_etal01}.  More
recently, \citet{sanderson_etal03} presented analysis of gas fractions
in a large sample of clusters spanning a wide range of masses.  The
resulting values of gas fraction depend on the assumptions made, but
overall the values are larger than in our simulations.  For instance
the mean gas fraction for their clusters is $0.13\pm 0.01$ or in units
of our universal baryon fraction $91\%$. Similarly to our simulated
clusters, the real clusters in their sample show gas fractions
monotonically increasing with cluster-centric radius.
\citet{sanderson_etal03} also found a trend of increasing gas fraction
with increasing cluster mass \citep[see
also][]{mohr_etal99,arnaud_evrard99} roughly consistent with the trend
for our sample of simulated clusters (see Figure~\ref{fig:fbgmv} and
eq.~[\ref{eq:fgas_fit}]). A caveat is that gas fractions in these
studies have been estimated either using the $\beta$-model fits to the
data or virial scaling relations rather than directly measuring gas
and total mass within $r_{500}$. 

We also note that \citet{bryan00} used observational data compiled
from the literature to argue that there is a trend of decreasing
stellar mass fractions with increasing cluster mass: $f_{\rm
  \ast}=0.042(T/10{\rm\, keV})^{-0.35}$, where $T$ is the
emission-weighted ICM temperature.  Given that $T\propto M^{2/3}$
according to the expected virial scaling, the corresponding scaling as
a function of cluster mass, $f_{\ast}\propto M^{-0.23}$, and 
its normalization are in good agreement with our simulations.

More recently, high sensitivity of the {\sl Chandra} and {\sl
  XMM-Newton} satellites allowed reliable measurements of temperature
profiles and direct estimates of gas fractions from the full
hydrostatic equilibrium analyses.  For instance, \citet{allen_etal04}
present gas fraction profiles at $r<r_{2500}$ for a sample of 26
massive clusters ($T_X>5$~keV) at different redshifts ($0<z<1$). They
quote gas fractions $f_{\rm gas}(<r_{2500})\approx 0.12$ for the
cosmology adopted in our study.  The corresponding gas fraction for
the massive cluster (CL1) in our study is $0.08$. This lower value may
be due to the fact that most of the clusters in the
\citet{allen_etal04} are more massive than CL1\footnote{There are some
  clusters in their sample that have $f_{\rm gas}(<r_{2500})<0.1$} or
may indicate overcooling in simulations. It may also be simply due to
the higher value of the universal baryon fraction in the real
Universe.  For example, if we use the value preferred by the WMAP data
\citep{spergel_etal03}, $\Omega_{\rm b}/\Omega_{\rm m}\simeq 0.17$,
the CL1 gas fraction in units of the universal value agrees with the
observed values in \citet{allen_etal04}.  A larger sample of massive
simulated clusters or comparison with a sample of lower mass clusters
is needed for a reliable conclusion.

The results presented in this paper have a number of implications for
the analyses that use clusters to obtain cosmological constraints.
First of all, our results show that cooling changes the cumulative
baryon and gas fractions appreciably even at the virial radius. Within
the inner regions of clusters ($r\lesssim r_{2500}$) the difference
between cooling and adiabatic simulations of the same cluster can be
as large as $\sim 20-40\%$. The effects are smaller for more massive
clusters or if the real clusters have lower fraction of condensed
baryons than clusters in our simulations. However, even for the
simulation in which stellar fraction is twice smaller
($f_{\ast}\approx 0.15$ in the re-simulation of CL3 with cooling
turned off at $z<2$) we find $20\%$ change in gas fraction and
$\approx 7\%$ change in the total baryon fraction within $r_{2500}$
compared to the adiabatic run. These differences are still large if
baryon fractions are to be used for precision cosmological constraints.  Our
analysis also shows that in simulations with cooling the total baryon
fraction can either be larger or smaller than the universal fraction
depending on the chosen radius and cluster mass. Therefore, there is
no single universal correction for converting the measured baryon
fraction to the universal value. These results suggest that care
should be taken in observational analyses to estimate baryon
fractions, and the systematic errors are likely mitigated if the
measurements are made at radii least affected by cooling and least
sensitive to the uncertain contribution of stars to the baryon budget.

Cooling and star formation decrease the fraction of hot intracluster
gas within the virial radius of the cluster. This decrease may affect
the Sunyaev-Zeldovich (SZ) fluxes and their relation to cluster mass.
Upcoming large SZ surveys will have a large fraction of objects with
masses of a few$\times 10^{14}\rm\ M_{\odot}$.  In such clusters star
formation can decrease the gas fraction by $\sim 20-40\%$ depending on
cluster mass.  The trend of gas fraction with mass will also modify
the slope of the relation between SZ flux and cluster mass. The trend
is weak but may need to be taken into account to ensure accurate
cosmological constraints.

Given the importance of these issues for the use of clusters as
cosmological probes, further effort from both theorists and observers
is needed to evaluate effects of cooling and to make robust
measurements of baryon fractions (gas$+$stars) in clusters.
Theoretically, it is important to simulate larger samples of clusters
over a wide range of masses and with high resolution and to
investigate the role of various processes (such as, for example,
thermal conduction and AGN feedback; e.g., \citeauthor{bruggen_etal05}
\citeyear{bruggen_etal05}) in suppressing cooling and star formation
in cluster cores, as well as processes such as helium segregation
which can affect cluster baryon fractions
\citep{qin_wu00,chuzhoy_loeb04}.  However, it is even more important
to first understand how large is the difference between the stellar
fractions in observed and simulated clusters.  If the fraction of the
diffuse intracluster stars is as large as $50\%$ \citep[][see however
\citeauthor{zibetti_etal05}
\citeyear{zibetti_etal05}]{gonzalez_etal04,lin_mohr04}, $f_{\ast}$ in
our simulations actually agrees with observations. Careful
observational calibration of stellar masses in clusters \cite[see,
e.g.,][]{voevodkin_etal02b,lin_etal04} using deep images in red or
infrared bands would go a long way in clarifying this issue.

\acknowledgements

We would like to thank Gustavo Yepes for providing us the gas fraction
profiles of the clusters simulated with the Gadget code and useful
discussions, and Volker Springel for making the entropy-conserving
version of Gadget available.  This project was supported by the
National Science Foundation (NSF) under grants No.  AST-0206216 and
AST-0239759, by NASA through grant NAG5-13274, and by the Kavli
Institute for Cosmological Physics at the University of Chicago. D.N.
is supported by the NASA Graduate Student Researchers Program and by
NASA LTSA grant NAG5--7986. AV is supported by the NASA grant
NAG5-9217 and contract NAS8-39073. The cosmological simulations used
in this study were performed on the IBM RS/6000 SP4 system ({\tt
copper}) at the National Center for Supercomputing Applications
(NCSA).

\bibliography{fbar}

\end{document}